\documentclass[final]{siamltex}

\usepackage{amsmath,amssymb,graphicx}
\usepackage[latin1]{inputenc}

\newcommand{\nc}{\newcommand}
\newcommand{\rnc}{\renewcommand}

\newcommand{\C}{{\mathbb C}}

\newcommand{\D}{{\mathbb D}}
\newcommand{\R}{{\mathbb R}}

\rnc{\Re}{\operatorname{Re}}
\rnc{\Im}{\operatorname{Im}}
\nc{\ol}{\overline}

\title{The perfect lens on a finite bandwidth}

\author{Øyvind Lind-Johansen\thanks{Department of Mathematical Sciences, Norwegian University of Science and Technology (NTNU), NO-7491 Trondheim, Norway, ({\tt lindjoha@stud.ntnu.no})},
\and Kristian Seip\thanks{Department of Mathematical Sciences, Norwegian University of Science and Technology (NTNU), NO-7491 Trondheim, Norway, ({\tt seip@math.ntnu.no}), supported by the Research Council of Norway grant 160192/V30.}
\and Johannes Skaar\thanks{Department of Electronics and Telecommunications, Norwegian University of Science and Technology (NTNU), NO-7491 Trondheim, Norway, ({\tt johannes.skaar@iet.ntnu.no})}}

\begin{document}

\maketitle

\begin{abstract}
The resolution associated with the so-called perfect lens of
thickness $d$ is\\ $-2\pi d/\ln(|\chi+2|/2)$. Here the
susceptibility $\chi$ is a Hermitian function in $H^2$ of the upper
half-plane, i.e., a $H^2$ function satisfying
$\chi(-\omega)=\overline{\chi(\omega)}$. An additional requirement
is that the imaginary part of $\chi$ be nonnegative for nonnegative
arguments. Given an interval $I$ on the positive half-axis, we
compute the distance in $L^\infty(I)$ from a negative constant to
this class of functions. This result gives a surprisingly simple and
explicit formula for the optimal resolution of the perfect lens on a
finite bandwidth.
\end{abstract}

\begin{keywords}
Metamaterials, negative refraction, perfect lens, dispersion, Hilbert transforms, $H^p$ spaces. \end{keywords}

\begin{AMS}
78A25, 
78A10, 
30D55. 
\end{AMS}

\pagestyle{myheadings}
\thispagestyle{plain}
\markboth{Ø. LIND-JOHANSEN, K. SEIP, AND J. SKAAR}{PERFECT LENS ON A FINITE BANDWIDTH}

\maketitle

\section{Introduction}

The resolution associated with imaging in conventional optics is of
the order a wavelength. This is a severe limitation in a number of
applications in nanoscience, e.g., in lithography, microscopy, and
spectroscopy. A remedy to this impasse---a so-called perfect
lens---was proposed by J.~B. Pendry \cite{pendry2000}. His idea was
to use metamaterials that allow for tailoring the dielectric
permittivity $\epsilon$ and the magnetic permeability $\mu$ by
structuring the medium at a length scale much smaller than a
wavelength \cite{pendry1996,pendry1999,smith}. This may lead to
negative refraction and restoration of the near-fields. The perfect
lens is a negative refraction metamaterial slab of a certain
thickness $d$. The metamaterial is considered linear, isotropic,
homogeneous, and without spatial dispersion. These ideal limits may
be approached by appropriate metamaterial designs. Note however that
certain implementations may inherently violate some of these
assumptions. For example, in a metal slab there is necessarily
spatial dispersion (nonlocality) of the dielectric response; this
limits the resolution to roughly 5 nm \cite{larkin2005}.

As the permittivity and permeability of the perfect lens are
negative, the material is necessarily dispersive \cite{veselago}.
Thus the perfect lens conditions $\epsilon=-1$ and $\mu=-1$ can only
be approached at a single frequency. Since most practical
applications involve a finite bandwidth, this fact limits the
performance of the perfect lens.

In the present work, we will quantify the finite bandwidth behavior.
The resolution at a single frequency is found to be
$r=-2\pi d/\ln(|\chi+2|/2)$, where $\chi=\epsilon-1$ or
$\chi=\mu-1$, depending on the incident polarization. Thus the
imaginary parts of $\epsilon$ and $\mu$ and the real parts'
deviation from $-1$ are both crucial for the resolution. The
formula for $r$ suggests an interesting mathematical problem:
Given a physically realizable susceptibility and an interval
$I=[a,b]$ on the positive angular frequency half-axis, find the
infimum of $\|\chi+2\|_{L^\infty(I)}$. Quite remarkably, this
problem can be solved explicitly, and as a result we obtain a simple
formula for the optimal resolution on a finite bandwidth.

\section{Resolution at a single frequency and on a finite bandwidth}
Let $\omega$ denote the (real) angular frequency. The
realizability criteria are causality, conjugate symmetry, and
passivity. They can be expressed as follows:
\begin{subequations}
\label{constraintschi}
\begin{align}
& \chi\in H^2,\label{KKchi}\\
& \chi(-\omega)=\overline{\chi(\omega)},\label{sym}\\
& \Im \chi(\omega)>0 \text{ for } \omega>0. \label{loss}
\end{align}
\end{subequations}
The causality criterion \eqref{KKchi} stating that $\chi$ belongs to
the Hardy space $H^2$ of the upper half-plane, means that the real
and imaginary parts of $\chi$ form a Hilbert transform pair
(Kramers--Kronig relations)
\cite{nussenzveig,landau_lifshitz_edcm}.\footnote{If the medium is
conducting at zero frequency, the electric susceptibility
$\epsilon-1$ is singular at $\omega=0$. Then \eqref{KKchi} is still
valid provided we set $\chi=\epsilon-1-i\sigma/\omega$, where
$\sigma>0$ is the zero frequency conductivity
\cite{landau_lifshitz_edcm}. In other words, such a medium will have
larger loss $\Im\epsilon$ for the same variation in $\Re\epsilon$.
With no loss of generality we can therefore exclude such media.} The
conjugate symmetry \eqref{sym} is a result of the fact that the
time-domain response function (inverse Fourier transform of $\chi$)
must be real \cite{landau_lifshitz_edcm}. Condition \eqref{loss} is
valid for all passive media, i.e., media in thermodynamic
equilibrium in the absence of the variable field
\cite{landau_lifshitz_edcm}.

\begin{figure}
\begin{center}
\includegraphics[width=6.5cm]{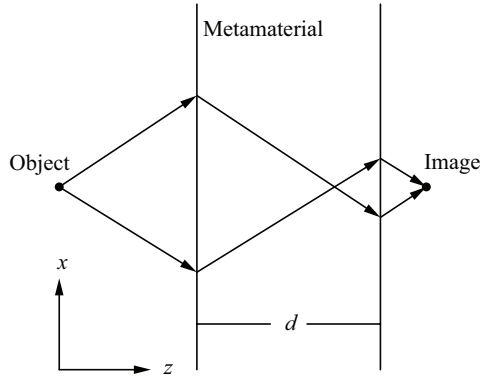}
\end{center}
\caption{The perfect lens: A metamaterial slab surrounded by vacuum. The distance from the object to the lens plus the distance from the lens to the image equals $d$. The arrows refer to the case when the metamaterial parameters $\epsilon$ and $\mu$ are close to $-1$.}
\label{fig:slab}
\end{figure}

The resolution as a function of $d$, $\epsilon$, and $\mu$ can be found by solving Maxwell's equations \cite{ramakrishna2002,nieto-vesperinas}. First we assume that the object to be imaged is one-dimensional. The slab has orientation orthogonal to the $z$-axis, the object varies along the $x$-direction, and the polarization of the magnetic field is taken to be along the $y$-axis, see Fig. \ref{fig:slab}. 
The object and the slab are surrounded by vacuum. For each spatial frequency $k_x$ of the object, we define the transmission coefficient $T$ as the ratio between the plane wave amplitudes at the image and the object. By matching tangential electric and magnetic fields at both surfaces, we find
\begin{equation}\label{tc1}
T=\frac{4\epsilon k_z k'_z e^{i(k_z-k'_z)d}}{(\epsilon k_z+k'_z)^2
e^{-2ik'_z d}-(\epsilon k_z-k'_z)^2},
\end{equation}
where $k_z^2=\omega^2/c^2-k_x^2$ and
$k_z'^{2}=\epsilon\mu\omega^2/c^2-k_x^2$. The sign of $k_z$ must be
chosen such that $\Im k_z\geq 0$, while the sign of $k_z'$ does not
matter in \eqref{tc1}. We are now interested in what happens for
large values of $k_x$, corresponding to near-fields that decay
exponentially in vacuum. We therefore assume that $k_x \gg\omega/c$.
With the additional assumptions that $|\epsilon+1|\ll 1$, $|\mu|$ at
most of the order $1$, and $d$ at most of the order a vacuum
wavelength ($\omega d/c\lesssim 1$), we obtain
\begin{equation}\label{tc2}
1/T\approx 1-\left[(1-\epsilon^2)-(\epsilon-\mu)
(\omega/ck_x)^2\right]^2\exp(2k_x d)/16,
\end{equation}
where we have used that \[\epsilon k_z+k'_z=(\epsilon^2
k_z^2-k_z'^{2})/(\epsilon
k_z-k'_z)=[(1-\epsilon^2)k_x^2+\epsilon(\epsilon-\mu)(\omega/c)^2]/(\epsilon
k_z-k'_z).\] We take the resolution $r$ to be the smallest value of
$2\pi/k_x$ such that the modulus of the second term on the right-hand side of \eqref{tc2} is equal to 1. This definition makes sense no matter what this term's phase happens to be, since the exponential factor $\exp(2k_xd)$ will force $|T|$ to decrease rapidly when $2\pi/k_x$ gets smaller than $r$. In general, the resolution becomes a nontrivial function of $\epsilon$ and $\mu$, but if
$|\epsilon-\mu|(\omega/ck_x)^2\ll 2|\epsilon+1|$, then
\begin{equation}\label{resolution1}
r = - \frac{2\pi d}{\ln{\frac{|\epsilon+1|}{2}}}.
\end{equation}
By the assumption $\omega/ck_x\ll 1$, the requirement
$|\epsilon-\mu|(\omega/ck_x)^2\ll 2|\epsilon+1|$ can be rewritten as
\begin{equation}\nonumber
|\epsilon-\mu|(\omega/c)d\lesssim -2|\epsilon+1|\ln\frac{|\epsilon+1|}{2}.
\end{equation}
Thus \eqref{resolution1} is always valid if $\epsilon=\mu$. Also, it
is valid provided the lens is sufficiently thin. Note that in the
latter case the resolution is independent of $\mu$. If we had chosen
the opposite polarization (i.e., electric field along the $y$-axis),
we would have arrived at exactly the same result only with the roles
of $\epsilon$ and $\mu$ interchanged. In other words, for a
one-dimensional object it is sufficient to have one of the
parameters $\epsilon$ and $\mu$ close to $-1$ with a small imaginary
part \cite{pendry2000}. For a two-dimensional object, both
polarizations are necessarily present; thus both $\epsilon$ and
$\mu$ should be close to $-1$.

The $L^\infty$-norm of the resolution restricted to the interval $I$
measures the poorest resolution in the corresponding frequency band.
In order to optimize the resolution, we should therefore minimize
the $L^\infty$ norm on $I$. In terms of the electric or magnetic
susceptibility $\chi=\epsilon-1$ or $\chi=\mu-1$, our task will
therefore be to compute the following distance:
\begin{equation}\nonumber
\inf \|\chi+2\|_{L^\infty(I)}.
\end{equation}

\section{The main result}

Before stating our main result, we introduce a few notational
conventions.

For $0<p<\infty$, the Hardy space $H^p$ of the upper half-plane
$\{\zeta=\omega+i\eta:\ \eta>0\}$ consists of those analytic
functions $f$ in this domain for which
\[ \|f\|_{H^p}^p=\sup_{\eta>0} \int_{-\infty}^\infty |f(\omega+i\eta)|^p d\omega <\infty;\]
$H^\infty$ is the space of bounded analytic functions. A function
$f$ in $H^p$ has nontangential boundary limits at almost every point
of the real axis, and the corresponding limit function, also denoted
$f$, is in $L^p=L^p(\R)$. Indeed, the $L^p$ norm of the boundary
limit function coincides with the $H^p$ norm introduced above. Thus
we may view $H^p$ as a subspace of $L^p$.

As already noted (see \eqref{KKchi}), we will require the following
symmetry condition: $f(-\omega)=\overline{f(\omega)}$. Functions $f$
satisfying this condition will be referred to as Hermitian
functions. We observe that Hermitian functions have even real parts
and odd imaginary parts.

The Hilbert transform of a function $u$ in $L^p$ ($1\le p<\infty$)
is defined as
\[ \tilde{u}(\omega)=\text{p.v.}\ \frac{1}{\pi} \int_{-\infty}^\infty
\frac{u(t)}{\omega-t} dt. \] It acts boundedly on $L^p$ for
$1<p<\infty$ and isometrically on $L^2$. If $u$ is a real-valued
function in $L^p$ for $1<p<\infty$, then $u+i\tilde{u}$ is in $H^p$,
and so the role of the Hilbert transform is to link the real and
imaginary parts of functions in $H^p$. We will only work with
Hermitian functions, and we will be interested in computing real
parts from imaginary parts. For this reason, it will be convenient
for us to consider the following Hilbert operator:
\[\mathcal{H}v(\omega)=\text{p.v.}\ \frac{1}{\pi}\int_{0}^\infty
v(t)\left(\frac{1}{t-\omega}+\frac{1}{t+\omega}\right) dt,\] acting
on functions in $L^p(\R^+)$. Provided $1<p<\infty$, the function
$\mathcal Hv+iv$ will then be in $H^p$, with the presumption that
$v$ is an odd function.

For a finite interval $I=[a,b]$ ($0<a<b$) set
\[ \Delta=\frac{b^2-a^2}{b^2+a^2}. \]

Our main theorem now reads as follows.

\begin{theorem}
\label{theorem1}

\[ \inf\left\{\|\mathcal{H}v+iv+2\|_{L^\infty(I)}: \ v\ge 0,\ v\in
L^2(\R^+)\right\}=\frac{2\Delta}{1+\sqrt{1-\Delta^2}}.\]
\end{theorem}

It will become clear that the infimum is not a minimum, but we may
extract from our proof explicit functions that bring us as close
as we wish to the infimum.

\section{Auxiliary results} The main result of \cite{SS05} will play a central role in
the proof of Theorem \ref{theorem1}. To state it, we define for each
real $\alpha$ the family of functions
\[ K_\alpha(I)=\{v\in L^2(\R^+):\ v(\omega)\ge 0 \
\text{for}  \  \omega>0, \mathcal{H} v(\omega)=\alpha \ \text{for} \ \omega\in
I\}.
\]
(Here and elsewhere we suppress the obvious ``almost everywhere''
provisions needed when considering pointwise restrictions.) We
think of functions in $K_\alpha(I)$, or more generally functions
in $L^2(\R^+)$, as the imaginary parts of Hermitian functions, and
we view them therefore as odd functions on $\R$.

We also need the function
\[ \sigma(\zeta)=\frac{1}{\sqrt{\zeta^2-b^2}\sqrt{\zeta^2-a^2}}. \]
It is taken to be positive for real arguments $w>b$ and is
analytic in the slit plane $\C\setminus([-b,-a]\cup[a,b])$. For
real arguments $a<|\zeta|<b$ we define $\sigma(\zeta)$ by extending it
continuously from the upper half-plane. Thus $\sigma(\zeta)$ takes
values on the negative imaginary half-axis when $\zeta$ is in $(a,b)$
and on the positive imaginary half-axis when $-\zeta$ is in $(a,b)$,
and otherwise it is real for real arguments.

Let us also associate with the interval $I$ the following Hilbert
operator:
\[\mathcal{H}_Iv(\omega)=\text{p.v.}\
\frac{1}{\pi}\int_{\R^+\setminus I}
v(t)\left(\frac{1}{t-\omega}+\frac{1}{t+\omega}\right) dt.\]

The main result of \cite{SS05} was the following parametrization
of $K_\alpha(I)$.

\begin{theorem}
\label{theorem2}
A nonnegative function $v$ in $L^2(\R^+)$ is in $K_\alpha(I)$ if
and only if the following three conditions hold:

\begin{equation} \label{one} \int_{\R^+\setminus I} v(t) |\sigma(t)| dt < \infty
\end{equation}
\begin{equation} \label{two} \frac{2}{\pi}\int_{\R^+\setminus I} tv(t) \sigma(t) dt =
\alpha \end{equation} \begin{equation} \label{three}
v(\omega)=\mathcal{H}_I\left(\sigma
v\right)(\omega)/|\sigma(\omega)|, \ \ \ \omega\in I.
\end{equation}

\end{theorem}

The integrability condition \eqref{one} is merely a slight growth
condition at the endpoints of $I$; we may write it more succinctly
as
\[
\int_0^a\left[v(a-t)+v(b+t)\right]\frac{dt}{\sqrt{t}} <\infty.
\]
This condition ensures that the integral in \eqref{two} and the
Hilbert transform appearing in \eqref{three} are both
well-defined.

At first sight, the theorem may not seem to give an explicit
parametrization of $K_\alpha(I)$. However, the Hilbert transform
appearing in \eqref{three} is given by
\[ \mathcal{H}_I\left(\sigma v\right)(\omega)=\frac{1}{\pi}\int_{\R^+\setminus I} v(t)
\sigma(t)\frac{2t}{t^2-\omega^2}  dt, \] and we observe that the
integrand on the right is nonnegative whenever $v(t)$ is
nonnegative. Hence $v(\omega)\ge 0$ for $\omega$ off $I$ implies $v(\omega)\ge 0$
for $\omega$ in $I$. This small miracle implies that $K_\alpha(I)$ is
parameterized by those nonnegative functions $v$ in
$L^2(\R^+\setminus I)$ for which \eqref{one} and \eqref{two} hold
and such that
\[ \int_I |\mathcal{H}_I\left(\sigma v\right)(\omega)|^2 |\sigma(\omega)|^{-2}
d\omega <\infty. \] By rephrasing this condition in more explicit
terms, we arrive at the following corollary \cite{SS05}.

\begin{corollary}
\label{corollary1} A nonnegative function $\nu$ in
$L^2(\R^+\setminus I)$ has an extension to a function in some class
$K_\alpha(I)$ if and only if the following condition holds:
\[ \int_0^a
\int_0^a\left[\nu(a-t)\nu(a-\tau)+\nu(b+t)\nu(b+\tau)\right]\frac{|\log(t+\tau)|}{\sqrt{t\tau}}
\ dtd\tau<\infty. \]
\end{corollary}
The difference between \eqref{one} and the condition above is the
logarithmic factor, which means that the condition of the
corollary is only a very slight strengthening of \eqref{one}. It
is clear that for instance boundedness of $v$ near the endpoints
of $I$ is more than enough.

We note that the integrand in \eqref{two} is negative to the left
of $I$ and positive to the right of $I$. This means that if
$\alpha$ is negative, then
\[ |\alpha|\le \frac{2}{\pi}\int_{0}^a tv(t) |\sigma(t)| dt, \]
with equality holding if $v$ vanishes to the right of $I$. It
follows that \begin{equation}\mathcal{H}_I\left(\sigma
v\right)(\omega)\ge \frac{1}{\pi}\int_0^a v(t)
\sigma(t)\frac{2t}{t^2-\omega^2} dt \ge \frac{|\alpha|}{\omega^2};
\label{optimal}
\end{equation}
we may come as close as we wish to this lower bound by choosing
any suitable $v$ supported by a small set sufficiently close to
$0$.

We remark that the results stated above can be proved by a method
similar to that given in the next section, the key ingredient being
a conformal map sending the ``two-sided'' segment $[a^2,b^2]$ to the
unit circle.

\section{Proof of Theorem \ref{theorem1}} In what follows, $H^\infty(\Omega)$
denotes the space of bounded analytic functions on $\Omega$ equipped
with the supremum norm, and $H^p(\D)$ stands for the $H^p$ spaces of
the open unit disk $\D$.

Let $v$ be an arbitrary function in $L^2(\R^+)$ such that $v\ge 0$
and
\[ \| \mathcal{H}v+iv \|_{L^\infty(I)}<\infty. \]
We will also assume\footnote{This assumption may seem unjustified.
However, we may first restrict attention to the smaller interval
$I_\varepsilon=[a+\varepsilon, b-\varepsilon]$ so that $v$ is bounded near the
endpoints of $I_\varepsilon$. Letting $\varepsilon\to 0$, we would then obtain
the same lower bound as we do with our a priori assumption.} that
$\nu=v|_{\R^+\setminus I}$ satisfies the condition of
Corollary~\ref{corollary1}. We know from Theorem \ref{theorem2} that
the extension of $\nu$ to a function in some class $K_\alpha(I)$ is
unique, and we may therefore use the notation $\nu$ for this
extension as well. If now $\varphi$ denotes an arbitrary bounded
function supported on $I$ with $\mathcal{H}\varphi$ also bounded, we
get the inequality
\[ \|\mathcal{H}v+iv+2 \|_{L^\infty(I)}\ge
\inf_\varphi
\|\mathcal{H}\varphi+i\varphi+i\nu+2+\alpha\|_{L^\infty(I)}.
\]
Setting \[\nu_0(t)=\begin{cases} \nu(t), & t\in I \\
                                    0, & t\in \R^+\setminus I,
                                    \end{cases} \]
we may write  \[ \|\mathcal{H}v+iv+2 \|_{L^\infty(I)}\ge
\inf_\varphi \|\mathcal{H}\varphi+i\varphi-\frac12(\mathcal{H}
\nu_0-i \nu_0)+2+\alpha\|_{L^\infty(I)}.
\]
The function $\mathcal{H}\varphi+i\varphi$ is the boundary limit
function of a bounded analytic function in the upper half-plane. In
fact, since the imaginary part has limit $0$ for every point in
$\R\setminus([-b,-a]\cup [a,b])$, this function extends by Schwarz
reflection to a bounded analytic function $f$ in $\C^*\setminus
([-b,-a]\cup [a,b])$, where $\C^*=\C\cup \{\infty\}.$ This function
$f$ satisfies $f(\ol{\zeta})=\ol{f(\zeta)}$ and $f(\infty)=0$. The function
$\mathcal{H} \nu_0+i \nu_0$ extends in the same fashion to a
function $g$ satisfying $g(\ol{\zeta})=\ol{g(\zeta)}$ and $g(\infty)=0$. Both these functions are in fact analytic in the variable $\xi=\zeta^2$; we write
\[ F(\xi)=f(\sqrt{\xi}) \ \ \text{and} \ \ G(\xi)=g(\sqrt{\xi}), \]
and then it follows that
\[ \inf_\varphi
\|\mathcal{H}\varphi+i\varphi-\frac12(\mathcal{H} \nu_0-i
\nu_0)+2+\alpha\|_{L^\infty(I)}=\inf_{F:\ F(\infty)=0}\|F+\frac12
\ol{G}+ 2+\alpha \|_{H^\infty(\C^*\setminus[a^2,b^2])}.\] The
remaining computation is most easily done if we first map
$\C^*\setminus[a^2,b^2]$ conformally onto the open unit disk $\D$,
say by the map
\[ w=w(\xi)=\frac{2}{b^2-a^2}\left(\xi-\frac{b^2+a^2}{2}+
\sqrt{\left(\xi-\frac{b^2+a^2}{2}\right)^2-\left(\frac{b^2-a^2}{2}\right)^2}\right),\]
where the square root is positive for positive arguments. We write
$\Gamma(w)=G(\xi(w))$ and obtain
\[ \inf_\varphi
\|\mathcal{H}\varphi+i\varphi-\frac12(\mathcal{H}\nu_0-i
\nu_0)+2+\alpha\|_{L^\infty(I)}=\inf_{F:\ F(0)=0}\|F+\frac12
\ol{\Gamma}+ 2+\alpha \|_{H^\infty(\D)}.\] Since $\Gamma(0)=F(0)=0$,
we have by orthogonality
\[ \|F+\frac12
\ol{\Gamma}+ 2+\alpha \|_{H^\infty(\D)}^2 \ge
\|\Gamma\|_{H^2(\D)}^2+(2+\alpha)^2. \] We may assume $\alpha\le
0$ since for $\alpha=0$ we may choose $\Gamma\equiv 0$. By
\eqref{three} and \eqref{optimal}, the expression on the
right-hand side is minimal if \[
\nu_0(\omega)=\frac{|\alpha|\sqrt{(b^2-\omega^2)(\omega^2-a^2)}}{\omega^2}.
\] For $|w|=1$, we therefore get
\[ \Im\Gamma(w)=\frac{|\alpha| \Im w}{\Delta^{-1}+\Re w}. \]
Hence
\[ \Gamma(w)= \frac{2\delta |\alpha| w}{1+\delta w},\]
where $\delta<1$ is determined by
the equation
\[\frac{2\delta}{1+\delta^2}=\Delta,\]
or in other words, \[
\delta=\Delta^{-1}-\sqrt{\Delta^{-2}-1}=\frac{\Delta}{1+\sqrt{1-\Delta^2}}.
\]
 It follows that
 \[ \|F+\frac12
\ol{\Gamma}+ 2+\alpha \|_{H^\infty(\D)}\ge
\left((2+\alpha)^2+\alpha^2 \frac{
\delta^2}{1-\delta^2}\right)^{1/2}\ge 2\delta;\] the minimum
on the right is obtained when
\[ \alpha= -2(1-\delta^2). \]

It remains for us to prove the remarkable fact that this lower bound
is in fact an infimum. To this end, observe that
\[ 2\delta^2+2(1-\delta^2)\delta \frac{w}{1+\delta
w}=2\delta \frac{w+\delta}{1+\delta w}. \] In other words,
the minimum is achieved if we choose $\alpha= -2(1-\delta^2)$,
$\varphi \equiv 0$, and
\[ \Gamma(w)=\frac{2\delta(1-\delta)^2 w}{1+\delta w}.\]
In view of \eqref{three} and \eqref{optimal}, we see that we can get
as close as we wish to the associated minimum by picking a function
$v\ge 0$ such that $v|_{\R^+\setminus I}$ is supported on a small
set close to $0$,
\begin{equation}\label{vweight}
\frac{2}{\pi}\int_{\R^+\setminus I} tv(t) \sigma(t) dt =
-2(1-\delta^2),
\end{equation}
and
\begin{equation}\label{vinI}
v(\omega)=\frac12 \mathcal{H}_I\left(\sigma v\right)(\omega)/|\sigma(\omega)|, \ \ \ \omega\in I. 
\end{equation}

\section{Discussion and conclusion}
\begin{figure}
\begin{center}
\includegraphics[height=7.5cm,width=8.1cm]{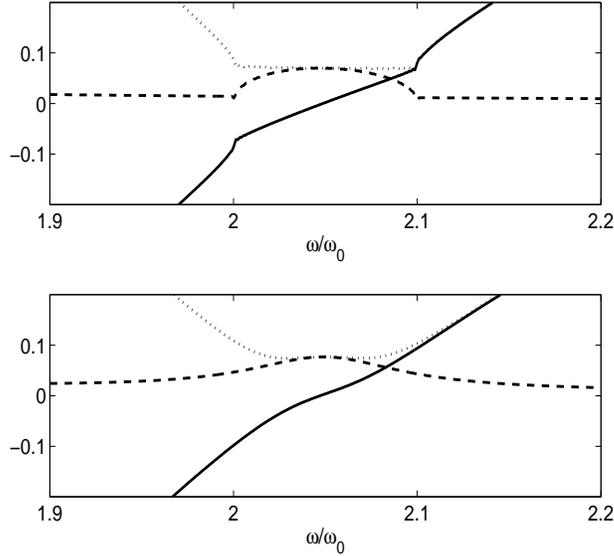}
\end{center}
\caption{The real parts (solid lines) and imaginary parts (dashed lines) of $\chi(\omega)+2$ for a near-optimal case (upper plot) and a Lorentzian approximation (lower plot). Also shown is the associated modulus $|\chi(\omega)+2|$ (dotted lines). Only the frequency range in the vicinity of $I$ is shown; in addition there is a strong resonance at $\omega_0$. The Hilbert transforms are computed using fast Fourier transforms. The parameters used are $a=2\omega_{0}$, $b=2.1\omega_{0}$, and $\gamma=0.1\omega_{0}$.}
\label{fig:optfunc}
\end{figure}

Theorem \ref{theorem1} together with \eqref{resolution1} gives the
optimal resolution as a function of bandwidth:
\begin{equation}\label{resolution2}
r = -\frac{2\pi d}{\ln{\frac{\Delta}{1+\sqrt{1-\Delta^2}}}}\approx -\frac{2\pi d}{\ln\frac{\Delta}{2}},
\end{equation}
where the approximation is valid for $\Delta\ll 1$. In this limit, $\Delta\approx(b-a)/b$ is the relative bandwidth. This optimal resolution is approached when the susceptibility $\chi(\omega)=u(\omega)+iv(\omega)$ contains a strong resonance at low frequencies, and a weak resonance centered at the relevant bandwidth. As an example, let $v(\omega)$ off $I$ be the imaginary part of a Lorentzian resonance function, i.e., $v(\omega)=\Im l(\omega)$ for $\omega\in \R^+\setminus I$, where 
\[ l(\omega)=\frac{L\omega_0^2}{\omega_0^2-\omega^2-i\omega\gamma}.\]
Here the resonance frequency $\omega_0$ and the bandwidth $\gamma$ are positive constants, while the strength $L$ is found from the requirement \eqref{vweight}. Computing $v(\omega)$ for $\omega\in I$ using \eqref{vinI}, and computing $u(\omega)=\mathcal{H}v(\omega)$, we obtain the upper plot in Fig. \ref{fig:optfunc}. Note that with this procedure, one may get arbitrarily close to the bound \eqref{resolution2} by choosing $\omega_0$ and $\gamma$ sufficiently small. Although realizable in principle, it might be difficult to fabricate a metamaterial with this near-optimal response. As an alternative, we can approximate the near-optimal susceptibility by letting $\chi(\omega)=u(\omega)+iv(\omega)$ be the superposition of two Lorentzian functions; one centered at $\omega_0$ and one centered at $(a+b)/2$. We let the first Lorentzian be equal to that in the former example. The bandwidth of the second Lorentzian is $b-a$, and we choose the strength such that $du((a+b)/2)/d\omega$ coincides with the corresponding value for the near-optimal case. The result is given in the lower plot in Fig. \ref{fig:optfunc}.

Since the distance from the object to the lens plus the distance
from the lens to the image equals $d$, there may be practical
reasons for not reducing $d$. If so, the resolution can only be
reduced by shrinking the operational bandwidth. Unfortunately, the
logarithmic dependence of $\Delta$ may require an unpractical, small
bandwidth.

It has been suggested to use a multilayer stack of alternating
negative index and positive index materials as the lens
\cite{ramakrishna2003}. This effectively reduces $d$ in
\eqref{resolution2}; however, then the distance from the object to
the lens plus the distance from the lens to the image equals the
thickness of each layer. In the limit when the layer thicknesses
approach zero, the resulting effective medium acts as a fiber-optic
bundle, but one that acts on the near field \cite{ramakrishna2003}.
To optimize the resolution and minimize aberrations, one again ends
up with the problem of minimizing $|\epsilon+1|$ and/or $|\mu+1|$. If aberrations can be tolerated, only $\Im\epsilon$ and/or $\Im\mu$ need to be minimized. This can be achieved on a finite bandwidth at the expense of some variation of $\Re\epsilon$ and/or $\Re\mu$ \cite{landau_lifshitz_edcm,SS06}.

We finally note that by simple scaling our result can be used to quantify the operational bandwidth of all components with desired permittivity $\epsilon_{\text{des}}$ and/or
permeability $\mu_{\text{des}}$ less than unity; with a certain tolerance of
$|\epsilon_{\text{des}}-\epsilon|$ and/or $|\mu_{\text{des}}-\mu|$. Thus, our result may for
instance prove useful for establishing the operational bandwidth of
invisibility cloaks \cite{pendry2006}. For components with
permittivity and permeability larger than or equal to unity, there
is no bandwidth limitation resulting from \eqref{constraintschi}.

\newpage
\bibliographystyle{siam}
\bibliography{bibfile.bib}

\def\cprime{$'$}
\begin{thebibliography}{10}

\bibitem{landau_lifshitz_edcm}
{\sc L.~D. Landau and E.~M. Lifshitz}, {\em Electrodynamics of continuous
  media}, Pergamon Press, New York and London, Chap. 9, 1960.

\bibitem{larkin2005}
{\sc I.~A. Larkin and M.~I. Stockman}, {\em Imperfect perfect lens}, Nano
  Letters, 5 (2005), pp.~339--343.

\bibitem{nieto-vesperinas}
{\sc M.~Nieto-Vesperinas}, {\em Problem of image superresolution with a
  negative-refractive-index slab}, J. Opt. Soc. Am. A, 21 (2004), pp.~491--498.

\bibitem{nussenzveig}
{\sc H.~M. Nussenzveig}, {\em Causality and dispersion relations}, Academic
  Press, New York and London, Chap. 1, 1972.

\bibitem{pendry2000}
{\sc J.~B. Pendry}, {\em Negative refraction makes a perfect lens}, Phys. Rev.
  Lett., 85 (2000), pp.~3966--3969.

\bibitem{pendry1999}
{\sc J.~B. Pendry, A.~J. Holden, D.~J. Robbins, and W.~J. Stewart}, {\em
  Magnetism from conductors and enhanced nonlinear phenomena}, IEEE Trans.
  Microwave Theory Tech., 47 (1999), pp.~2075--2084.

\bibitem{pendry1996}
{\sc J.~B. Pendry, A.~J. Holden, W.~J. Stewart, and I.~Youngs}, {\em Extremely
  low frequency plasmons in metallic mesostructures}, Phys. Rev. Lett., 76
  (1996), pp.~4773--4776.

\bibitem{pendry2006}
{\sc J.~B. Pendry, D.~Schurig, and D.~R. Smith}, {\em Controlling
  electromagnetic fields}, Science, 312 (2006), pp.~1780--1782.

\bibitem{ramakrishna2002}
{\sc S.~A. Ramakrishna, J.~B. Pendry, D.~Schurig, D.~R. Smith, and S.~Schultz},
  {\em The asymmetric lossy near-perfect lens}, J. Mod. Optics, 49 (2002),
  pp.~1747--1762.

\bibitem{ramakrishna2003}
{\sc S.~A. Ramakrishna, J.~B. Pendry, M.~C.~K. Wiltshire, and W.~J. Stewart},
  {\em Imaging the near field}, J. Mod. Optics, 50 (2003), pp.~1419--1430.

\bibitem{SS05}
{\sc K.~Seip and J.~Skaar}, {\em An extremal problem related to negative
  refraction}, Skr. K. Nor. Videns. Selsk., 3 (2005), pp.~1--8; arXiv.org/math.CV/0506620.


\bibitem{SS06}
{\sc J.~Skaar and K.~Seip}, {\em Bounds for the refractive indices of
  metamaterials}, Journal of Physics D: Applied Physics, 39 (2006),
  pp.~1226--1229.

\bibitem{smith}
{\sc D.~R. Smith, W.~J. Padilla, D.~C. Vier, S.~C. Nemat-Nasser, and
  S.~Schultz}, {\em Composite medium with simultaneously negative permeability
  and permittivity}, Phys. Rev. Lett., 84 (2000), p.~4184.

\bibitem{veselago}
{\sc V.~G. Veselago}, {\em The electrodynamics of substances with
  simultaneously negative values of $\epsilon$ and $\mu$}, Soviet Physics
  Uspekhi, 10 (1968), p.~509.

\end{thebibliography}

\end{document}